\documentclass[showpacs,pre,aps,twocolumn]{revtex4-1}

\pdfoutput=1
\usepackage{amsmath,amsfonts,amssymb,bm,graphicx,color}

\usepackage{hyperref}
\usepackage{color}

\renewcommand{\L}{\mathcal{L}}
\newcommand{\W}{\mathcal{W}}
\newcommand{\R}{\mathcal{R}}

\newcommand{\A}{\mathbb{A}}
\newcommand{\M}{\mathbb{M}}
\newcommand{\K}{\mathbb{K}}
\newcommand{\Q}{\mathbb{Q}}

\newcommand{\F}{\mathcal{F}}

\newcommand{\er}[1]{Eq.~\eqref{#1}}
\newcommand{\ers}[2]{Eqs.~(\ref{#1}-\ref{#2})}
\newcommand{\era}[2]{Eqs.~(\ref{#1}) and (\ref{#2})}
\newcommand{\eraa}[3]{Eqs.~(\ref{#1}), (\ref{#2}) and (\ref{#3})}

\newcommand{\fs}{\langle - |}

\begin{document}

\title{Simple bounds on fluctuations and uncertainty relations for first-passage times of counting observables}

\author{Juan P. Garrahan}
\affiliation{School of Physics and Astronomy}
\affiliation{Centre for the Mathematics and Theoretical Physics of Quantum Non-equilibrium Systems,
University of Nottingham, Nottingham NG7 2RD, UK}

\date{\today}

\begin{abstract}
Recent large deviation results have provided general lower bounds for the fluctuations of time-integrated currents in the
steady state of stochastic systems.  A corollary are so-called thermodynamic uncertainty relations connecting precision of estimation to average dissipation. Here we consider this problem but for counting observables, i.e., trajectory observables which, in contrast to currents, are non-negative and non-decreasing in time (and possibly symmetric under time reversal).  In the steady state, their fluctuations to all orders are bound from below by a Conway-Maxwell-Poisson distribution dependent only on the averages of the observable and of the dynamical activity.  We show how to obtain the corresponding bounds for first-passage times (times when a certain value of the counting variable is first reached) and their uncertainty relations.  Just like entropy production does for currents, dynamical activity controls the bounds on fluctuations of counting observables. 
 
\end{abstract}
\maketitle

\section{Introduction}

In this note we try to connect three recent developments in the theory of stochastic systems.  The first are general {\em bounds on the fluctuations of time-integrated currents} \cite{Gingrich2016,Pietzonka2016,Pietzonka2016b,Gingrich2016b}.  Obtained by means ``Level 2.5'' \cite{Maes2008,Bertini2012,Bertini2015} dynamical large deviation methods \cite{Dembo1998,Ruelle2004,Gaspard2005,Lecomte2007,Touchette2009}, these results stipulate general lower bounds for fluctuations at any order of all empirical currents in the stationary state of a stochastic process \cite{Gingrich2016,Pietzonka2016,Pietzonka2016b,Gingrich2016b}.  A corolary are {\em thermodynamic uncertainty relations} \cite{Barato2015,Barato2015b,Polettini2016} connecting the estimation error of time-integrated currents to overall dissipation.

The second development are {\em fluctuation relations for first-passage times} (FPTs) \cite{Roldan2015,Saito2016,Neri2016}, similar to those of more standard observables such as work or entropy production.  From these an uncertainty relation connecting dissipation to the time needed to determine the direction of time can be derived \cite{Roldan2015}.  These results indicate a relation between the fluctuations of observables in dynamics over a fixed time, with fluctuations in stopping times.  

The third development is {\em trajectory ensemble equivalence} \cite{Chetrite2013,Chetrite2015,Budini2014,Kiukas2015} between ensembles of long trajectories subject to different constraints.  For example, for long times, the ensemble of trajectories conditioned on a fixed value of a time-integrated quantity is equivalent to that conditioned only on its average \cite{Chetrite2013,Chetrite2015} (cf.\ microcanonical/canonical equivalence of equilibrium ensembles \cite{Peliti2011}).  Similarly, the ensemble of trajectories of fixed total time and fluctuating number of jumps is equivalent to that of fixed number of jumps but fluctuating time \cite{Budini2014,Kiukas2015} (cf.\ fixed volume and fixed pressure static ensembles \cite{Peliti2011}).  

The works in Refs.\  \cite{Gingrich2016,Pietzonka2016,Pietzonka2016b,Gingrich2016b} and \cite{Barato2015,Barato2015b,Polettini2016,Roldan2015,Saito2016,Neri2016} focus on trajectory observables asymmetric under time reversal, such as empirical currents, which can be positive or negative and can increase and decrease with time.  
Here we consider instead trajectory observables which are always non-negative and strictly non-decreasing with time.  We call these {\em counting observables}.  An example is the total number of configuration changes in a trajectory, 
or {\em dynamical activity} \cite{Lecomte2007,Garrahan2007,Maes2008}.  Here show that from the bounds to the rate functions of counting observables, via trajectory ensemble equivalence, we can derive the corresponding bounds for arbitrary fluctuations of FPTs.  

After introducing the basics of dynamical large deviations, in Sec.~III we show that the rate functions of counting observables are bounded from above by a Conway-Maxwell-Poisson distribution (a generalisation of the Poisson distribution that allows for non-Poissonian number fluctuations \cite{Shmueli2005}).  The corresponding bound for the cumulant generating function was first found in Ref.~\cite{Pietzonka2016} (called ``exponential bound''); here we rederive it straightforwardly via Level 2.5 large deviations, cf.\ \cite{Gingrich2016}.  In Sec.\ IV we consider the large deviations of FPTs and establish the correspondence between the FTP and observable generating functions. This allows, in Sec.\ V, to derive a general bound on FPT rate functions from the large deviations of the observable distributions.  From these bounds FPT uncertainty relations follow.  An important observation is that the bounds on fluctuations of a counting observable and its FPTs are controlled by the average dynamical activity, in analogy to the role played by the entropy production in the case of currents \cite{Gingrich2016,Barato2015}.  We hope these results will add to the growing body of work applying large deviation ideas and methods to the study of non-equilibrium dynamics in classical and quantum stochastic systems, see e.g., 
\cite{Lebowitz1999,Evans2004,Merolle2005,Appert2008,Garrahan2009,Baiesi2009,Hedges2009,Kurchan2009,Esposito2009,Garrahan2010,Giardina2011,Budini2011,Speck2011,Speck2012b,Bodineau2012,Flindt2013,Weber2013,Espigares2013,Mey2014,Vaikuntanathan2014,Weber2015,Jack2015,Ueda2015,DeBacco2015,Szavits2015,Jack2015b,Horssen2015,Pigeon2015,Verley2016,Bonanca2016,Nemoto2016,Speck2016,Jack2016,Gherardini2016,Deffner2016,Karevski2016}.

\section{Stochastic dynamics and large deviations of counting observables}

We consider systems evolving as continuous time Markov chains~\cite{Gardiner2004}, with master equation,
\begin{align}
\partial_{t} P_t(x) = \sum_{x,y \neq x} W_{yx} P_t(y) - \sum_x R_x P_t(x) , 
\label{ME1}
\end{align}
where $P_t(x)$ is the probability being in configuration $x$ at time $t$, $W_{xy}$ the transition rate from $x$ to $y$, and $R_x = \sum_{y \neq x} W_{xy}$ the escape rate from $x$. In operator form the master equation reads,   
\begin{equation}
\partial_{t} |P_t \rangle = \L |P_t \rangle ,
\label{ME2}
\end{equation}
with probability vector $|P_t\rangle = \sum_x P_t(x) | x \rangle$, where $\{ | x \rangle \}$ is an orthonormal configuration basis. The master operator is,
\begin{align}
\L &= \W - \R = \sum_{x,y \neq x} W_{xy} |y \rangle \langle x|
- \sum_x R_x |x \rangle \langle x|,
\label{L}
\end{align}
where $\W$ and $\R$ indicate the off-diagonal and diagonal parts of $\L$, respectively.  This dynamics is realised by stochastic trajectories, such as $\omega = ( x_{0} \to x_{t_1} \to \ldots \to x_{t_K} )$. This trajectory has $K$ jumps, with the jump between configurations $x_{t_{i-1}}$ and $x_{t_{i}}$ occurring at time $t_{i}$, with $0 \leq t_1 \leq \cdots t_K \leq t$, and no jump between $t_K$ and $t$. We denote by $\pi_t(\omega)$ the probability of $\omega$ within the ensemble of trajectories of total time $t$.

Properties of the dynamics are encoded in trajectory observables, i.e., functions of the whole trajectory, $\A(\omega)$, which are extensive in time.  Examples include time-integrated currents or dynamical activities.  Time-exentisivity implies that at long times their probabilities and moment generating functions have large deviation forms \cite{Dembo1998,Ruelle2004,Gaspard2005,Lecomte2007,Touchette2009},
\begin{align}
P_t(A) &= \sum_{\omega} \delta \left[ A - \A(\omega) \right] \pi_t(\omega)
\approx e^{-t \varphi(A/t)} ,
\label{PA} 
\\
Z_t(s) &= \sum_{\omega} e^{-s \A(\omega)} \pi_t(\omega) \approx e^{t \theta(s)} ,
\label{Z}
\end{align}
where the rate function $\varphi(a)$ and the scaled cumulant generating function $\theta(s)$ are related by a Legendre transform \cite{Dembo1998,Ruelle2004,Gaspard2005,Lecomte2007,Touchette2009}, 
\begin{equation}
\varphi(a) = -\min_s \left[ \theta(s) + s \, a \right] .
\label{LT}
\end{equation}

In what follows we focus on trajectory observables defined in terms of the jumps in a trajectory, 
\begin{equation}
\A(\omega) = \sum_{xy} \alpha_{xy} \Q_{xy}(\omega), 
\label{A}
\end{equation}
where $\Q_{xy}(\omega)$ is the number of jumps from $x$ to $y$ in trajectory $\omega$.  We will assume all $\alpha_{xy} \geq 0$.  This means that $\A(\omega)$ is non-negative and non-decreasing with time.  We call $\A(\omega)$ a {\em counting observable} as it counts the number of certain kinds of jumps in the trajectory.  Furthermore, when $\alpha_{xy} = \alpha_{yx}$ these observables are symmetric under time-reversal, in contrast to time-integrated currents which are antisymmetric (and therefore neither necessarily positive nor non-decreasing with time).  An important example of a counting observable is the total number of jumps or dynamical activity \cite{Lecomte2007,Garrahan2007,Maes2008},
\begin{equation}
\K(\omega) = \sum_{xy} \Q_{xy}(\omega).
\label{K}
\end{equation}

For observables such as \er{A} the moment generating function \er{Z} can be written as 
\begin{equation}
Z_t(s) = \fs e^{t \L_s} | x_0 \rangle ,
\label{ZLs}
\end{equation}
where $\L_s$ is the tilted operator \cite{Dembo1998,Ruelle2004,Gaspard2005,Lecomte2007,Touchette2009},
\begin{align}
\L_s = \W_s - \R
= \sum_{x,y \neq x} e^{- \alpha_{xy} s} W_{xy} |y \rangle \langle x|
- \R ,
\label{Ls}
\end{align}
and $\fs = \sum_{x} \langle x|$. The function $\theta(s)$ is then given by the largest eigenvalue of $\L_s$.

\section{Level 2.5 and fluctuation bounds}

The computation of large deviation functions as the ones in \era{PA}{Z} for arbitrary observables and dynamics is difficult in general.  There is however one case where the rate function can be written down explicitly \cite{Maes2008,Bertini2012,Bertini2015}. 

If we denote by $\M_x(\omega)$ the total residence time in configuration $x$ throughout trajectory $\omega$, then $t^{-1} \M_x(\omega)$ is the {\em empirical measure}.  Similarly, from the number of jumps $\Q_{xy}(\omega)$ we can define the empirical {\em flux}, $t^{-1} \Q_{xy}(\omega)$.  Since $\M(\omega)$ and $\Q(\omega)$ are extensive observables, their probability obeys a large deviation principle at long times \cite{Dembo1998,Ruelle2004,Gaspard2005,Lecomte2007,Touchette2009},
\begin{align}
P_t(q,m) = & \sum_\omega
\pi_t(\omega)
\prod_x \delta \left[ m_x - t^{-1} \M_x(\omega) \right]
\label{pqm}
\\
& ~~ \times
\prod_{xy} \delta \left[ q_{xy} - t^{-1} \Q_{xy}(\omega) \right]
\approx
e^{-t I(q,m)} .
\nonumber
\end{align}
The rate function $I(q,m)$ has an explicit form in the stationary state dynamics of \er{ME2}, known as ``level 2.5'' of large deviations \cite{Maes2008,Bertini2012,Bertini2015},  
\begin{equation}
I(q,m) = \sum_{xy} q_{xy} \left[ \ln\left( \frac{q_{xy}}{m_x W_{xy}} \right) - 1 \right] + \sum_x m_x R_x ,
\label{l25}
\end{equation}
where $m$ and $q$ must obey the probability conserving conditions, 
\begin{equation}
\sum_x m_x = 1 \; , \;\;\; \sum_y q_{xy} = \sum_y q_{yx} .
\label{cons}
\end{equation}
This rate function is minimised (its minimum value being zero) when $m$ and $q$ take the stationary average values
\begin{equation}
m_x = \rho_x \; , \;\; q_{xy} = \rho_x W_{xy} ,
\label{l25min}
\end{equation}
where $\rho_x$ is the stationary distribution, $\L | \rho \rangle = 0$.  The rate function for a trajectory observable such as \er{A} can then be obtained by {\em contraction} \cite{Dembo1998,Ruelle2004,Gaspard2005,Lecomte2007,Touchette2009},
\begin{equation}
\varphi(a) = \min_{q,m ~:~ a = \alpha \cdot q} I(q,m) ,
\label{min}
\end{equation}
where $\alpha \cdot q = \sum_{xy} \alpha_{xy} q_{xy}$ and $a=A/t$. 

An upper bound for $\varphi(a)$ can be obtained following the procedure of Ref.\ \cite{Gingrich2016}.  From \er{min}, any pair of empirical measure $m$ and flux $q$ that satisfies \er{cons} and has $a = \sum_{xy} \alpha_{xy} q_{xy}$ will give an upper bound to $\varphi(a)$.  A convenient and simple choice is,
\begin{equation}
m^*_x = \rho_x \; , \;\; 
q^*_{xy} = \frac{a}{\langle a \rangle} \rho_x W_{xy} ,
\label{mqs}
\end{equation}
where $\langle a \rangle = \sum_{xy} \alpha_{xy} \rho_x W_{xy}$.  We then get, with $I_*(a) = I(q^*,m^*)$,
\begin{equation}
\varphi(a) \leq I_*(a) = \frac{\langle k \rangle}{\langle a \rangle} \left[ \ln \left( \frac{a}{\langle a \rangle} \right) - \left( a - \langle a \rangle \right) \right] ,
\label{ba}
\end{equation}
where $\langle k \rangle = \sum_{xy} \rho_x W_{xy} = \sum_{x} \rho_x R_{x} $ is the average dynamical activity (per unit time).  The rate function on the right side of \er{ba} is that of a Conway-Maxwell-Poisson (CMP) distribution \cite{Shmueli2005}, a generalisation of the Poisson distribution for a counting variable with non-Poissonian number fluctuations.  

From the Legendre transform \er{LT}, the upper bound \er{ba} also implies a lower bound for the scaled cumulant generating function $\theta(s)$,
\begin{equation}
\theta(s) \geq \theta_*(s) = \langle k \rangle 
\left[ \exp \left( - s \frac{\langle a \rangle}{\langle k \rangle} \right) - 1 \right] .
\label{bs}
\end{equation}
The expression on the right is the scaled cumulant generating function of a CMP distribution.  This last result was first derived in Ref.\ \cite{Pietzonka2016} in a slightly different manner.

Figure 1 illustrates the bounds \era{ba}{bs} for the elementary example of a two-level system.  The exact rate function $\varphi(a)$ and the upper bound $I_*(a)$ have the same minimum at $\langle a \rangle$, but the fluctuations of $a$ are larger than those given by $I_*(a)$ for all $a$.  The exact cumulant generating function $\theta(s)$ and its lower bound $\theta_*(s)$ have the same slope at $s=0$, but $\theta_*(s)$ has derivatives which are smaller in magnitude to all orders that those of $\theta(s)$, again indicating that the CMP approximation provides lower bounds for the size of fluctuations of $a$.  

As occurs with the analogous bounds on time-integrated currents \cite{Gingrich2016,Pietzonka2016,Pietzonka2016b,Gingrich2016b}, an immediate consequence of the bounds on the rate function or cumulant generating function are the {\em thermodynamic uncertainty relations} \cite{Barato2015,Barato2015b,Polettini2016}.  From \er{ba} or \er{bs} we get a lower bound for the variance of the observable in terms of its average and the average activity (cf.\ \cite{Pietzonka2016}) 
\begin{equation}
{\rm var}(a) = \frac{\theta''(0)}{t} \geq \frac{\theta_*''(0)}{t} = \frac{\langle a \rangle^2}{\langle k \rangle \, t} .
\label{bvar} 
\end{equation}
This in turn provides an upper bound for precision of estimation of the observable $A$ in terms of the signal-to-noise ratio (i.e.\ inverse of the error),
\begin{equation}
{\rm SNR}(A) = \frac{\langle A \rangle}{\sqrt{{\rm var}(A)}} \leq \sqrt{{\langle K \rangle}} \, ,
\label{bsnr}
\end{equation}
where ${\langle K \rangle} = t {\langle k \rangle}$.  Just like in the case of integrated currents \cite{Barato2015,Barato2015b,Polettini2016}, where there is an unavoidable tradeoff between precision and dissipation, the uncertainty in the estimation of a counting observable is bounded generically by the overall average activity in the process.

\begin{figure}[t!]
\begin{center}
\includegraphics[width=\columnwidth]{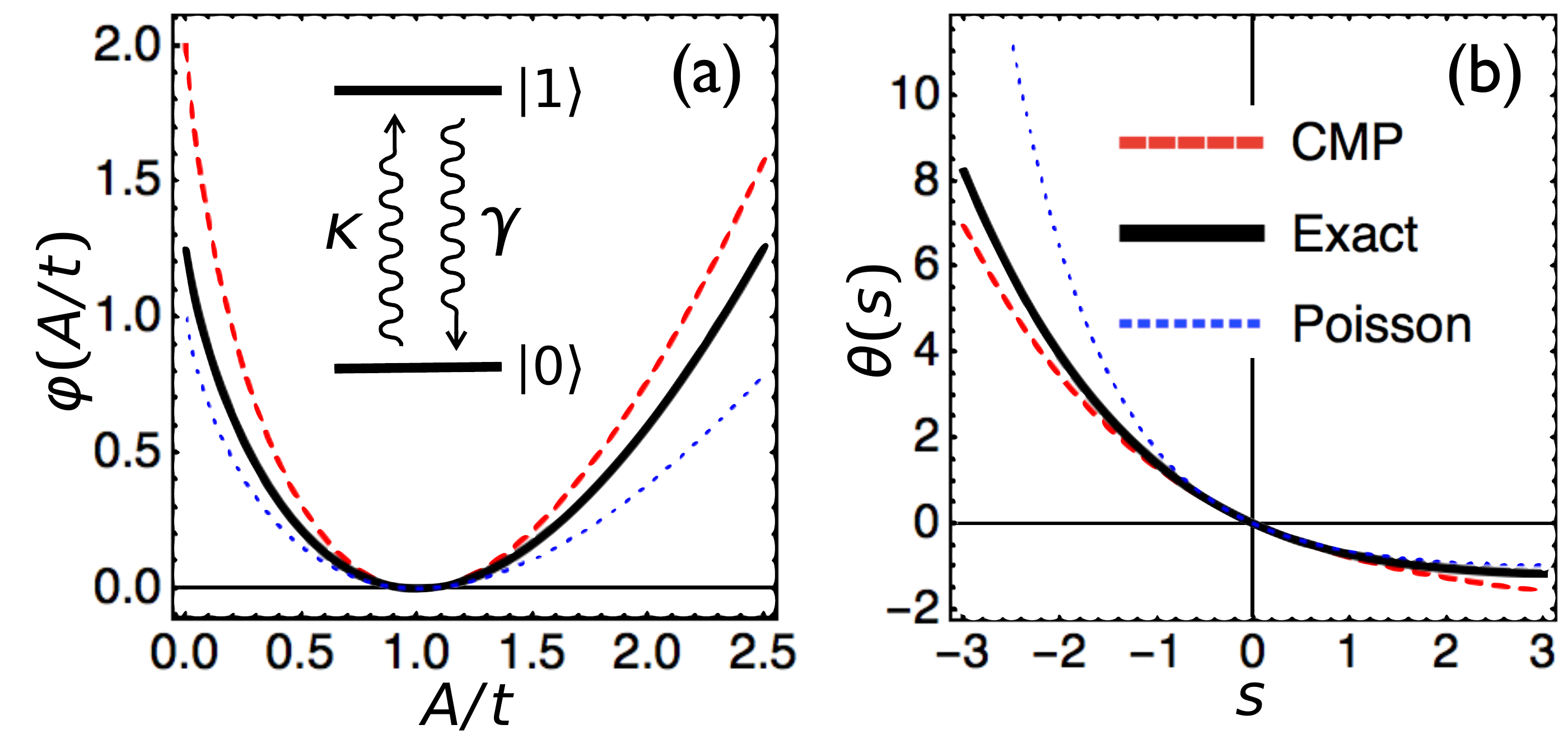}
\caption{Bounds on observable fluctuations for a 2-level system.  Transition rates are $W_{10}=\gamma$ and $W_{01}=\kappa$.  We consider as observable $A$ the total number of $1 \to 0$ jumps.  In the stationary state $\langle a \rangle =
\langle A \rangle/t = \gamma \kappa / (\gamma + \kappa)$.  The average activity per unit time is $\langle k \rangle = 2 \langle a \rangle$.  Panel (a) shows the rate function $\varphi(A/t)$ (full/black) for $\gamma=5$ and $\kappa=1.25$.   The rate function is bounded from above everywhere by a CMP rate function, \er{ba} (dashed/red).  We also show for comparison a Poisson rate function with mean $\langle a \rangle$ (dotted/blue).  Panel (b) shows the corresponding scaled cumulant generating function $\theta(s) = \frac{1}{2} \left[ \sqrt{(\gamma -\kappa)^2+4 \gamma  \kappa  e^{-s}}-(\gamma+\kappa)\right]$ (full/black).  It is monotonic in $s$ since $A \geq 0$, and is bounded from below, \er{bs}, by 
$\theta_*(s) = \frac{2 \gamma \kappa}{\gamma + \kappa} ( e^{-s/2} -1 )$ (dashed/red).  For the case $\kappa=\gamma$ the bounds become exact in this simple model.
}
\end{center}
\end{figure}

\section{Large deviations of first-passage time distributions}

We consider now the statistics of first-passage times (FPT) (also called {\em stopping times}), the times at which a certain trajectory observable first reaches a threshold value.  This implies a change of focus from ensembles of trajectories of total fixed time to ensembles of trajectories of fluctuating overall time \cite{Bolhuis2008,Budini2014,Harris2016}.  Recently, distributions of FPT associated with entropy production have been shown to obey fluctuation relations \cite{Roldan2015,Saito2016,Neri2016} reminiscent of those of current-like observables.  This suggests a duality between observable and FPT statistics, which in turn is connected to the equivalence between fixed time and fluctuating time trajectory ensembles, see e.g.~\cite{Budini2014,Kiukas2015}.  

We focus on stopping times for counting observables as defined in \er{A}.  For simplicity we assume that the coefficients $\alpha_{xy}$ are either 0 or 1, so that $\A(\omega)$ counts a subset of all possible jumps in a trajectory and takes integer values. (These assumptions can be relaxed at the expense of slightly more involved expressions without changing the essence of the results.) 

Lets consider the structure of trajectories associated with FPT events for a fixed value $A$ of the observable $\A(\omega)$.  Such a trajectory will have $A$ jumps for which $\alpha_{xy}=1$, occurring at times $0 \leq t_1 \leq \cdots t_{A-1} \leq t_A=\tau$ with $\tau$ being the FPT through $\A(\omega)=A$.  In between these jumps the evolution will be one where only jumps with $\alpha_{xy}=0$ occur.  The weight of this trajectory is related to the amplitude of a matrix product state \cite{Garrahan2016},
\begin{align}
\langle y | \,
\tilde{\W} \,
e^{(t_A-t_{A-1}) \L_\infty} \,
\cdots
\tilde{\W} \,
e^{(t_2-t_1) \L_\infty} \,
\tilde{\W} \,
e^{t_1 \L_\infty} \,
| x \rangle .
\label{mps}
\end{align}
This expression is the weight of all trajectories starting in $x$ and ending in $y$, after $A$ jumps that contribute to the observable, occurring at the specified times $t_i$ ($i=1,\ldots,A$), and with an arbitrary number of the other jumps.  Here $\L_\infty$ is the tilted operator \er{Ls} at $s \to \infty$, so that all transitions associated to $\A(\omega)$ are suppressed.  The factors $e^{\Delta t \L_\infty}$ encode dynamics which do not contribute to increasing the observable and which occur between the times $t_i$. The operator
\begin{equation}
\tilde{\W} = \L - \L_\infty ,
\label{tW}
\end{equation} 
includes all the transitions that increase $\A(\omega)$ by one unit, and \er{mps} has $A$ insertions of $\tilde{\W}$.  Integrating \er{mps} over intermediate times and summing over the final configuration formally yields the FPT distribution, 
\begin{equation}
F_{x}(\tau | A) = \int_{0 \leq t_1 \cdots \leq \tau}
\langle - | \,
\tilde{\W} \,
e^{(\tau-t_{A-1}) \L_\infty} \,
\cdots
\tilde{\W} \,
e^{t_1 \L_\infty} \,
| x \rangle .
\nonumber
\end{equation}
This expression simplifies via a Laplace transform,\begin{equation}
\hat{F}_{x}(\mu | A) = \int_0^\infty dt \, e^{-\mu \tau} \, F_{x}(\tau | A) = \langle - | \F_\mu^A | x \rangle ,
\label{F1}
\end{equation}
where the transfer operator reads
\begin{equation}
\F_\mu = \tilde{\W} \, \left( \mu - \L_\infty \right)^{-1} .
\label{F2}
\end{equation}
When $A$ is large, $A \gg 1$, the Laplace transformed FTP distribution has a large deviation form,
\begin{equation}
\hat{F}_{x}(\mu | A) \approx e^{A g(\mu)} ,
\label{FLD}
\end{equation}
where $e^{g(\mu)}$ is the largest eigenvalue of $\F_\mu$.  Note the similarities between \ers{F1}{FLD} and \ers{Z}{Ls}.

The eigenvalues of $\F_\mu$ and $\L_s$ are directly related.  From \eraa{Ls}{tW}{F2} we can write,
\begin{equation}
e^{-s} \F_\mu = \left( \L_s - \mu \right) \left( \mu - \L_\infty \right)^{-1} + 1 .
\label{feq}
\end{equation}
Consider now a row vector $\langle l |$ which is a left eigenvector both of $\F_\mu$ and $\L_s$, with eigenvalue $e^{g(\mu)}$ and $\theta(s)$, respectively.  Multiplying \er{feq} by $\langle l |$ and rearranging we get
\begin{equation}
\left(e^{-s+g(\mu)} - 1\right) \langle l |
= 
\left[ \theta(s) - \mu \right] \langle l | \left( \mu - \L_\infty \right)^{-1} .
\label{feq2}
\end{equation}
We see that for $\langle l |$ to be a simultaneous eigenvector of $\F_\mu$ and $\L_s$ we need to have $g(\mu)=s$ and $\theta(s)=\mu$.  That is, $g$ is the functional inverse of $\theta$ and vice-versa,
\begin{equation}
\theta(s) = g^{-1}(s), \;\;\; 
g(\mu) = \theta^{-1}(\mu) .
\label{tg}
\end{equation}
For the case where the counting observable is the dynamical activity, \er{K}, the analysis above is that of ``$x$-ensemble'' of Ref.~\cite{Budini2014}, i.e., the ensemble of trajectories of fixed total number of jumps but fluctuating time.

For the general problem of the FPTs for arbitrary counting observables, \ers{F1}{F2} coincide with the FPT distributions first obtained in Ref.~\cite{Saito2016} in a different way.  The derivation in Ref.~\cite{Saito2016} proceeds in the standard manner used for example in the proof of FPT distributions for diffusion processes \cite{Gardiner2004}. It relates the probability of having accumulated $A$ up to time $t$, to the  probability of reaching $A$ at time $\tau \leq t$ for the first time followed by no increment in $A$ from $\tau$ to $t$,
\begin{equation}
P_t(A|x) = \sum_y \int_0^t d\tau P_{t-\tau}(0|y) F_{xy}(\tau|A) , 
\label{SD1}
\end{equation}
where $P_t(A|x)$ is \er{PA} with the initial condition made explicit, and $F_{xy}(\tau|A)$ refers to the FPT distribution for time $\tau$ and final configuration $y$.  If we transform from $A$ to $s$, cf.~\eraa{PA}{Z}{ZLs}, we can rewrite \er{SD1} as matrix elements of
\begin{equation}
e^{t \L_s} = \int_0^t d\tau e^{(t-\tau) \L_\infty} \hat{\F}_s(\tau) , 
\label{SD2}
\end{equation}
where $\langle y | \hat{\F}_s(\tau) |x \rangle = \sum_A e^{-s A} F_{xy}(\tau|A)$.  After a Laplace transform and rearraging we get,
\begin{equation}
\hat{\F}_{s\mu} = \left( \mu - \L_\infty \right) \left( \mu - \L_s \right)^{-1} . 
\label{SD3}
\end{equation}
This last expression is the same as that in Ref.~\cite{Saito2016} after a discrete Laplace transform from $A$ to $s$.  We can invert the $A \to s$ transformation as follows.  The l.h.s.\ of \er{SD3} is,
\begin{equation}
\hat{\F}_{s\mu} = \sum_{A=0}^\infty e^{-s A} \hat{\F}_\mu(A) , 
\label{SD4}
\end{equation}
while the r.h.s.\ can be rewritten as, 
\begin{align}
\left( \mu - \L_\infty \right) & \left( \mu - \L_s \right)^{-1} 
\nonumber
\\ & =
\left[ 1 - e^{-s} \tilde{\W} \left( \mu - \L_\infty \right)^{-1} \right]^{-1} 
\nonumber
\\ & =
\sum_{A=0}^\infty e^{-s A} 
\left[ \tilde{\W} \left( \mu - \L_\infty \right)^{-1} \right]^{A} .
\label{SD5}
\end{align}
Equating \era{SD4}{SD5} term by term we get that 
\begin{equation}
\hat{\F}_\mu(A) = \F_\mu^A ,
\label{SD6}
\end{equation}
with $\F_\mu$ given by \er{F2}, showing that our derivation is equivalent to that of Ref.~\cite{Saito2016}.  The advantage of expressing the FPT distribution in terms of its generating function \er{F2} as we have done here is that it allows for a direct extraction of its large deviation function, see  \era{FLD}{tg}, giving access to the full statistics of FPTs in the limit of large $A$.

\begin{figure}[t!]
\begin{center}
\includegraphics[width=\columnwidth]{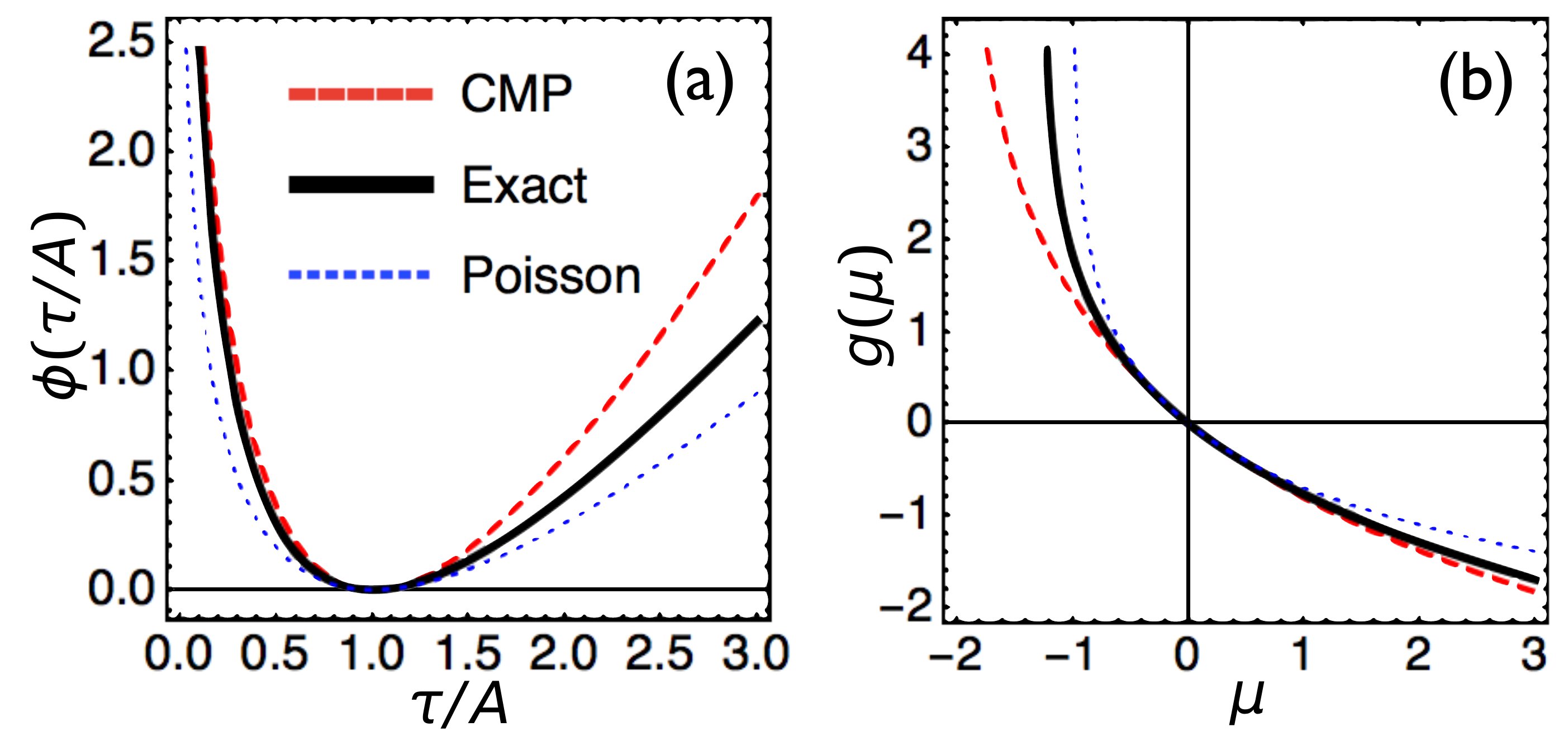}
\caption{Bounds on first-passage time fluctuations for the 2-level system of Fig.~1.  The FPT $\tau$ is defined as the time when a total $A$ of up/down jumps $1 \to 0$ is reached. In the stationary state $\langle \tau \rangle = A / \langle a \rangle = A (\gamma + \kappa) / (\gamma \kappa)$. Panel (a) shows the rate function $\phi(\tau/A)$ (full/black) for $\gamma=5$ and $\kappa=1.25$, and assuming the initial state is $0$. The rate function is bounded from above everywhere by $\phi_*(\tau/A)$, \er{bf} (dashed/red).  We also show for comparison the FPT rate function of a Poisson process with the same mean (dotted/blue).  
Panel (b) shows the FPT scaled cumulant generating function 
$g(\mu) = \ln(\gamma \kappa)-\ln[(\gamma+\mu)(\kappa+\mu)]$ 
(full/black).  It is bounded from below, \er{bg}, by 
$g_*(\mu)$ (dashed/red).
}
\end{center}
\end{figure}

\section{Bounds on FPT distributions}

Equations (\ref{F1}-\ref{tg}) establish a connection between the statistics of a counting observable, at fixed overall time, and the statistics of the FPT for a fixed value of said observable.  This connection is due to the equivalence \cite{Budini2014,Kiukas2015} between the ensemble of trajectories of fixed time, but where the observable is allowed to fluctuate (in a manner controlled by the field $s$ conjugate to the observable), with the ensemble of fixed observable but where the time extension of trajectories is allowed to fluctuate (in a manner controlled by the field $\mu$ conjugate to time).  This equivalence holds in the limit of large observable/time, where the relation between the controlling fields is given by \er{tg}.  We can now use the results of Sec.~III on bounds on observable fluctuations to infer the corresponding bounds on FPT fluctuations. 

The bound \er{bs} on the cumulant generating function of $A$ provides an lower bound to the FPT scaled cumulant generating function $g(\mu)$ through \er{tg}.  Inverting $\theta_*$ in \er{bs} we get 
\begin{equation}
g(\mu) \geq g_*(\mu) = - \frac{\langle k \rangle}{\langle a \rangle} \ln \left( \frac{\mu}{\langle k \rangle} + 1 \right) .
\label{bg}
\end{equation}
For large $A$ the FPT distribution also has a large deviation form, 
\begin{equation}
F(\tau | A) \approx e^{-A \phi(\tau/A)} ,
\label{Ftau}
\end{equation}
where $\phi(\tau/A)$ is obtained from $g(\mu)$ by a Legendre transform similar to \er{LT}.  From \er{bg} we then obtain an upper bound for the FPT rate function,
\begin{align}
\phi(\tau/A) \leq & ~ \phi_*(\tau/A) =
\label{bf}
\\
&
~~ - \frac{\langle k \rangle}{\langle a \rangle} \left[ 
\ln \left( \frac{\tau \langle a \rangle}{A} \right) 
- \left( \frac{\tau \langle a \rangle}{A} - 1 \right)
\right] .
\nonumber
\end{align}
Figure 2 illustrates the upper bound of the FPT rate function, \er{bf}, and the lower bound of the FPT cumulant generating function, \er{bg}, for the same 2-level model of Fig.~1. 

The bound function $\phi_*(\tau/A)$ has its minimum at the exact value of the average FPT,
\begin{equation}
\langle \tau \rangle_A = \frac{A}{\langle a \rangle} ,
\label{ta}
\end{equation}
where $\langle \cdot \rangle_A$ indicates average in the FPT ensemble of fixed $A$.  That the average FPT is given by the inverse of the observable per unit time follows immediately from \er{tg}.  The second derivative of $\phi_*(\tau/A)$ at its minimum provides a lower bound for the variance of the FPT.  From \er{bf}, or alternatively \er{bg}, we get, 
\begin{equation}
\frac{{\rm var}(\tau)}{A} = g''(0) \geq g_*''(0) = \frac{1}{\langle a \rangle \langle k \rangle} .
\label{tvar} 
\end{equation}
This in turn gives a bound on the precision with which one can estimate the FPT,
\begin{equation}
{\rm SNR}(\tau) = \frac{\langle \tau \rangle_A}{\sqrt{{\rm var}(\tau)}} \leq \sqrt{{\langle K_A \rangle}} \, .
\label{bsnrt}
\end{equation}
where ${\langle K_A \rangle} = \langle \tau \rangle_A \langle k \rangle$.  As for case of the uncertainty for the observable, \er{bsnr}, the precision of estimation of the FPT is limited by the total average activity, in this case for trajectories of length $t = \langle \tau \rangle_A$.

\section{Conclusions}

We have discussed general bounds on fluctuations of counting observables, hopefully complementing the more detailed recent results on current fluctuations 
\cite{Gingrich2016,Pietzonka2016,Pietzonka2016b,Gingrich2016b}.  While empirical currents are the natural trajectory observables to consider in driven problems \cite{Gingrich2016,Pietzonka2016,Pietzonka2016b,Gingrich2016b,Maes2008,Lebowitz1999,Evans2004,Appert2008,Espigares2013,Jack2015,Karevski2016}, counting observables such as the dynamical activity are central quantities for systems with complex equilibrium dynamics, such as glass formers \cite{Merolle2005,Garrahan2007,Garrahan2009,Hedges2009,Speck2012b}.  (And even for driven systems it is revealing to study the dynamical phase behaviour in terms of both empirical currents and activities, see e.g.\ \cite{Appert2008,Speck2011,Jack2015,Karevski2016}.)  

The bounds are a  straightforward consequence of the Level 2.5 large deviation \cite{Maes2008,Bertini2012,Bertini2015} description, \er{l25}, which provides an explicit (and useful) minimisation principle for rate functions.  But as remarked in \cite{Gingrich2016b}, these bounds may be more or less descriptive depending on whether they are tight or loose, which in turn depends on how good the variational ansatz is.  As observed in \cite{Pietzonka2016}, the ansatz \er{mqs} is akin to a mean-field approximation that homogenises the connections between states.  For any counting observable which is a subset of the overall activity the rate function is bound by a CMP distribution with sub-Poissonian number fluctuations, see \era{ba}{bs}.  For the elementary example of Fig.\ 1 the bound is tight, but more complex problems of interest often display large (that is, super-Poissonian) number fluctuations \cite{Merolle2005,Garrahan2007,Hedges2009,Garrahan2010,Speck2012b,Weber2013}.  It would be interesting to find alternative yet simple variational ansatzes that can capture such strong fluctuation behaviour.  Nevertheless, there are still important consequences that follow even from these simple bounds.  An immediate one is that the dynamical activity cannot be sub-Poissonian, which in turn implies an exponential in time complexity for the efficient sampling of trajectories conditioned on it, cf.~\cite{Jack2015}.  

We have also shown how to obtain related general bounds on the distributions of first-passage times.  Again this complements for counting observables, and generalises, recent results on FPTs for current-like quantities  \cite{Roldan2015,Saito2016,Neri2016}.  We did this by exploiting the correspondence between the large deviation functions of observables and those of FPTs, \ers{FLD}{tg}.  Note that this correspondence works for observables which are non-decreasing in time.  For these, the zero increment probability $P_{t}(0|y)$, \er{SD1}, is directly related to the tilted operator $\L_\infty$, leading to the ensemble correspondence, \ers{FLD}{tg}.  For currents, however, a zero observable does not imply the absence of jumps that contribute to the observable (only that their contribution adds up to zero), and the correspondence breaks down (or at least we have not been able to relate the corresponding cumulant generating functions in that case).  Just like in the case of activities, the FPTs are bounded by the distribution of times of a CMP process, \era{bg}{bf}, as illustrated in Fig.~2.

As occurs for currents \cite{Barato2015,Barato2015b,Polettini2016}, the bounds to rate functions give rise to thermodynamic uncertainty relations constraining the precision of estimation of both observables and FPTs, \era{bsnr}{bsnrt}.  For empirical currents, which are time-asymmetric, precision is limited by the average entropy produced \cite{Barato2015,Barato2015b,Polettini2016}.  In turn, for counting observables and their FPTs, the corresponding limit is set by the average dynamical activity, suggesting that this quantity might play as important a role in the dynamics as the overall dissipation.

\begin{acknowledgements}
This work was supported by EPSRC Grant No.~EP/M014266/1.
\end{acknowledgements}

%
%
%
%
%
%
%
%
%
%

\bibliography{bounds}

\begin{thebibliography}{65}%
\makeatletter
\providecommand \@ifxundefined [1]{%
 \@ifx{#1\undefined}
}%
\providecommand \@ifnum [1]{%
 \ifnum #1\expandafter \@firstoftwo
 \else \expandafter \@secondoftwo
 \fi
}%
\providecommand \@ifx [1]{%
 \ifx #1\expandafter \@firstoftwo
 \else \expandafter \@secondoftwo
 \fi
}%
\providecommand \natexlab [1]{#1}%
\providecommand \enquote  [1]{``#1''}%
\providecommand \bibnamefont  [1]{#1}%
\providecommand \bibfnamefont [1]{#1}%
\providecommand \citenamefont [1]{#1}%
\providecommand \href@noop [0]{\@secondoftwo}%
\providecommand \href [0]{\begingroup \@sanitize@url \@href}%
\providecommand \@href[1]{\@@startlink{#1}\@@href}%
\providecommand \@@href[1]{\endgroup#1\@@endlink}%
\providecommand \@sanitize@url [0]{\catcode `\\12\catcode `\$12\catcode
  `\&12\catcode `\#12\catcode `\^12\catcode `\_12\catcode `\%12\relax}%
\providecommand \@@startlink[1]{}%
\providecommand \@@endlink[0]{}%
\providecommand \url  [0]{\begingroup\@sanitize@url \@url }%
\providecommand \@url [1]{\endgroup\@href {#1}{\urlprefix }}%
\providecommand \urlprefix  [0]{URL }%
\providecommand \Eprint [0]{\href }%
\providecommand \doibase [0]{http://dx.doi.org/}%
\providecommand \selectlanguage [0]{\@gobble}%
\providecommand \bibinfo  [0]{\@secondoftwo}%
\providecommand \bibfield  [0]{\@secondoftwo}%
\providecommand \translation [1]{[#1]}%
\providecommand \BibitemOpen [0]{}%
\providecommand \bibitemStop [0]{}%
\providecommand \bibitemNoStop [0]{.\EOS\space}%
\providecommand \EOS [0]{\spacefactor3000\relax}%
\providecommand \BibitemShut  [1]{\csname bibitem#1\endcsname}%
\let\auto@bib@innerbib\@empty
\bibitem [{\citenamefont {Gingrich}\ \emph {et~al.}(2016)\citenamefont
  {Gingrich}, \citenamefont {Horowitz}, \citenamefont {Perunov},\ and\
  \citenamefont {England}}]{Gingrich2016}%
  \BibitemOpen
  \bibfield  {author} {\bibinfo {author} {\bibfnamefont {T.~R.}\ \bibnamefont
  {Gingrich}}, \bibinfo {author} {\bibfnamefont {J.~M.}\ \bibnamefont
  {Horowitz}}, \bibinfo {author} {\bibfnamefont {N.}~\bibnamefont {Perunov}}, \
  and\ \bibinfo {author} {\bibfnamefont {J.}~\bibnamefont {England}},\
  }\href@noop {} {\bibfield  {journal} {\bibinfo  {journal} {Phys. Rev. Lett.}\
  }\textbf {\bibinfo {volume} {116}},\ \bibinfo {pages} {120601} (\bibinfo
  {year} {2016})}\BibitemShut {NoStop}%
\bibitem [{\citenamefont {Pietzonka}\ \emph
  {et~al.}(2016{\natexlab{a}})\citenamefont {Pietzonka}, \citenamefont
  {Barato},\ and\ \citenamefont {Seifert}}]{Pietzonka2016}%
  \BibitemOpen
  \bibfield  {author} {\bibinfo {author} {\bibfnamefont {P.}~\bibnamefont
  {Pietzonka}}, \bibinfo {author} {\bibfnamefont {A.~C.}\ \bibnamefont
  {Barato}}, \ and\ \bibinfo {author} {\bibfnamefont {U.}~\bibnamefont
  {Seifert}},\ }\href@noop {} {\bibfield  {journal} {\bibinfo  {journal} {Phys.
  Rev. E}\ }\textbf {\bibinfo {volume} {93}},\ \bibinfo {pages} {052145}
  (\bibinfo {year} {2016}{\natexlab{a}})}\BibitemShut {NoStop}%
\bibitem [{\citenamefont {Pietzonka}\ \emph
  {et~al.}(2016{\natexlab{b}})\citenamefont {Pietzonka}, \citenamefont
  {Barato},\ and\ \citenamefont {Seifert}}]{Pietzonka2016b}%
  \BibitemOpen
  \bibfield  {author} {\bibinfo {author} {\bibfnamefont {P.}~\bibnamefont
  {Pietzonka}}, \bibinfo {author} {\bibfnamefont {A.~C.}\ \bibnamefont
  {Barato}}, \ and\ \bibinfo {author} {\bibfnamefont {U.}~\bibnamefont
  {Seifert}},\ }\href@noop {} {\bibfield  {journal} {\bibinfo  {journal} {J.
  Phys. A}\ }\textbf {\bibinfo {volume} {49}},\ \bibinfo {pages} {34LT01}
  (\bibinfo {year} {2016}{\natexlab{b}})}\BibitemShut {NoStop}%
\bibitem [{\citenamefont {Gingrich}\ \emph {et~al.}()\citenamefont {Gingrich},
  \citenamefont {Rotskoff},\ and\ \citenamefont {Horowitz}}]{Gingrich2016b}%
  \BibitemOpen
  \bibfield  {author} {\bibinfo {author} {\bibfnamefont {T.~R.}\ \bibnamefont
  {Gingrich}}, \bibinfo {author} {\bibfnamefont {G.~M.}\ \bibnamefont
  {Rotskoff}}, \ and\ \bibinfo {author} {\bibfnamefont {J.~M.}\ \bibnamefont
  {Horowitz}},\ }\href@noop {} {\bibinfo  {journal} {arXiv:1609.07131}\
  }\BibitemShut {NoStop}%
\bibitem [{\citenamefont {Maes}\ and\ \citenamefont
  {Netocny}(2008)}]{Maes2008}%
  \BibitemOpen
\bibfield  {journal} {  }\bibfield  {author} {\bibinfo {author} {\bibfnamefont
  {C.}~\bibnamefont {Maes}}\ and\ \bibinfo {author} {\bibfnamefont
  {K.}~\bibnamefont {Netocny}},\ }\href@noop {} {\bibfield  {journal} {\bibinfo
   {journal} {Europhys. Lett.}\ }\textbf {\bibinfo {volume} {82}},\ \bibinfo
  {pages} {30003} (\bibinfo {year} {2008})}\BibitemShut {NoStop}%
\bibitem [{\citenamefont {Bertini}\ \emph {et~al.}()\citenamefont {Bertini},
  \citenamefont {Faggionato},\ and\ \citenamefont {Gabrielli}}]{Bertini2012}%
  \BibitemOpen
  \bibfield  {author} {\bibinfo {author} {\bibfnamefont {L.}~\bibnamefont
  {Bertini}}, \bibinfo {author} {\bibfnamefont {A.}~\bibnamefont {Faggionato}},
  \ and\ \bibinfo {author} {\bibfnamefont {D.}~\bibnamefont {Gabrielli}},\
  }\href@noop {} {\bibinfo  {journal} {arXiv:1212.6908}\ }\BibitemShut
  {NoStop}%
\bibitem [{\citenamefont {Bertini}\ \emph {et~al.}(2015)\citenamefont
  {Bertini}, \citenamefont {Faggionato},\ and\ \citenamefont
  {Gabrielli}}]{Bertini2015}%
  \BibitemOpen
\bibfield  {journal} {  }\bibfield  {author} {\bibinfo {author} {\bibfnamefont
  {L.}~\bibnamefont {Bertini}}, \bibinfo {author} {\bibfnamefont
  {A.}~\bibnamefont {Faggionato}}, \ and\ \bibinfo {author} {\bibfnamefont
  {D.}~\bibnamefont {Gabrielli}},\ }\href@noop {} {\bibfield  {journal}
  {\bibinfo  {journal} {Stoc. Proc. Appl.}\ }\textbf {\bibinfo {volume}
  {125}},\ \bibinfo {pages} {2786} (\bibinfo {year} {2015})}\BibitemShut
  {NoStop}%
\bibitem [{\citenamefont {Dembo}\ and\ \citenamefont
  {Zeitouni}(1998)}]{Dembo1998}%
  \BibitemOpen
  \bibfield  {author} {\bibinfo {author} {\bibfnamefont {A.}~\bibnamefont
  {Dembo}}\ and\ \bibinfo {author} {\bibfnamefont {O.}~\bibnamefont
  {Zeitouni}},\ }\href@noop {} {\emph {\bibinfo {title} {Large Deviation
  Techniques and Applications}}},\ \bibinfo {edition} {2nd}\ ed.\ (\bibinfo
  {publisher} {Springer},\ \bibinfo {year} {1998})\BibitemShut {NoStop}%
\bibitem [{\citenamefont {Ruelle}(2004)}]{Ruelle2004}%
  \BibitemOpen
  \bibfield  {author} {\bibinfo {author} {\bibfnamefont {D.}~\bibnamefont
  {Ruelle}},\ }\href@noop {} {\emph {\bibinfo {title} {Thermodynamic
  formalism}}}\ (\bibinfo  {publisher} {Cambridge University Press},\ \bibinfo
  {year} {2004})\BibitemShut {NoStop}%
\bibitem [{\citenamefont {Gaspard}(2005)}]{Gaspard2005}%
  \BibitemOpen
  \bibfield  {author} {\bibinfo {author} {\bibfnamefont {P.}~\bibnamefont
  {Gaspard}},\ }\href@noop {} {\emph {\bibinfo {title} {Chaos, Scattering and
  Statistical Mechanics}}}\ (\bibinfo  {publisher} {Cambridge University
  Press},\ \bibinfo {year} {2005})\BibitemShut {NoStop}%
\bibitem [{\citenamefont {Lecomte}\ \emph {et~al.}(2007)\citenamefont
  {Lecomte}, \citenamefont {Appert-Rolland},\ and\ \citenamefont {van
  Wijland}}]{Lecomte2007}%
  \BibitemOpen
  \bibfield  {author} {\bibinfo {author} {\bibfnamefont {V.}~\bibnamefont
  {Lecomte}}, \bibinfo {author} {\bibfnamefont {C.}~\bibnamefont
  {Appert-Rolland}}, \ and\ \bibinfo {author} {\bibfnamefont {F.}~\bibnamefont
  {van Wijland}},\ }\href@noop {} {\bibfield  {journal} {\bibinfo  {journal}
  {J. Stat. Phys.}\ }\textbf {\bibinfo {volume} {127}},\ \bibinfo {pages} {51}
  (\bibinfo {year} {2007})}\BibitemShut {NoStop}%
\bibitem [{\citenamefont {Touchette}(2009)}]{Touchette2009}%
  \BibitemOpen
  \bibfield  {author} {\bibinfo {author} {\bibfnamefont {H.}~\bibnamefont
  {Touchette}},\ }\href@noop {} {\bibfield  {journal} {\bibinfo  {journal}
  {Phys. Rep.}\ }\textbf {\bibinfo {volume} {478}},\ \bibinfo {pages} {1}
  (\bibinfo {year} {2009})}\BibitemShut {NoStop}%
\bibitem [{\citenamefont {Barato}\ and\ \citenamefont
  {Seifert}(2015{\natexlab{a}})}]{Barato2015}%
  \BibitemOpen
  \bibfield  {author} {\bibinfo {author} {\bibfnamefont {A.~C.}\ \bibnamefont
  {Barato}}\ and\ \bibinfo {author} {\bibfnamefont {U.}~\bibnamefont
  {Seifert}},\ }\href@noop {} {\bibfield  {journal} {\bibinfo  {journal} {Phys.
  Rev. Lett.}\ }\textbf {\bibinfo {volume} {114}},\ \bibinfo {pages} {158101}
  (\bibinfo {year} {2015}{\natexlab{a}})}\BibitemShut {NoStop}%
\bibitem [{\citenamefont {Barato}\ and\ \citenamefont
  {Seifert}(2015{\natexlab{b}})}]{Barato2015b}%
  \BibitemOpen
  \bibfield  {author} {\bibinfo {author} {\bibfnamefont {A.~C.}\ \bibnamefont
  {Barato}}\ and\ \bibinfo {author} {\bibfnamefont {U.}~\bibnamefont
  {Seifert}},\ }\href@noop {} {\bibfield  {journal} {\bibinfo  {journal} {J.
  Phys. Chem. B}\ }\textbf {\bibinfo {volume} {119}},\ \bibinfo {pages} {6555}
  (\bibinfo {year} {2015}{\natexlab{b}})}\BibitemShut {NoStop}%
\bibitem [{\citenamefont {Polettini}\ \emph {et~al.}()\citenamefont
  {Polettini}, \citenamefont {Lazarescu},\ and\ \citenamefont
  {Esposito}}]{Polettini2016}%
  \BibitemOpen
  \bibfield  {author} {\bibinfo {author} {\bibfnamefont {M.}~\bibnamefont
  {Polettini}}, \bibinfo {author} {\bibfnamefont {A.}~\bibnamefont
  {Lazarescu}}, \ and\ \bibinfo {author} {\bibfnamefont {M.}~\bibnamefont
  {Esposito}},\ }\href@noop {} {\bibinfo  {journal} {arXiv:1605.09692}\
  }\BibitemShut {NoStop}%
\bibitem [{\citenamefont {Rold\'an}\ \emph {et~al.}(2015)\citenamefont
  {Rold\'an}, \citenamefont {Neri}, \citenamefont {D\"orpinghaus},
  \citenamefont {Meyr},\ and\ \citenamefont {J\"ulicher}}]{Roldan2015}%
  \BibitemOpen
\bibfield  {journal} {  }\bibfield  {author} {\bibinfo {author} {\bibfnamefont
  {E.}~\bibnamefont {Rold\'an}}, \bibinfo {author} {\bibfnamefont
  {I.}~\bibnamefont {Neri}}, \bibinfo {author} {\bibfnamefont {M.}~\bibnamefont
  {D\"orpinghaus}}, \bibinfo {author} {\bibfnamefont {H.}~\bibnamefont {Meyr}},
  \ and\ \bibinfo {author} {\bibfnamefont {F.}~\bibnamefont {J\"ulicher}},\
  }\href@noop {} {\bibfield  {journal} {\bibinfo  {journal} {Phys. Rev. Lett.}\
  }\textbf {\bibinfo {volume} {115}},\ \bibinfo {pages} {250602} (\bibinfo
  {year} {2015})}\BibitemShut {NoStop}%
\bibitem [{\citenamefont {Saito}\ and\ \citenamefont {Dhar}(2016)}]{Saito2016}%
  \BibitemOpen
  \bibfield  {author} {\bibinfo {author} {\bibfnamefont {K.}~\bibnamefont
  {Saito}}\ and\ \bibinfo {author} {\bibfnamefont {A.}~\bibnamefont {Dhar}},\
  }\href@noop {} {\bibfield  {journal} {\bibinfo  {journal} {EPL}\ }\textbf
  {\bibinfo {volume} {114}},\ \bibinfo {pages} {50004} (\bibinfo {year}
  {2016})}\BibitemShut {NoStop}%
\bibitem [{\citenamefont {Neri}\ \emph {et~al.}()\citenamefont {Neri},
  \citenamefont {Rold{\'a}n},\ and\ \citenamefont {J{\"u}licher}}]{Neri2016}%
  \BibitemOpen
  \bibfield  {author} {\bibinfo {author} {\bibfnamefont {I.}~\bibnamefont
  {Neri}}, \bibinfo {author} {\bibfnamefont {{\'E}.}~\bibnamefont
  {Rold{\'a}n}}, \ and\ \bibinfo {author} {\bibfnamefont {F.}~\bibnamefont
  {J{\"u}licher}},\ }\href@noop {} {\bibinfo  {journal} {arXiv:1604.04159}\
  }\BibitemShut {NoStop}%
\bibitem [{\citenamefont {Chetrite}\ and\ \citenamefont
  {Touchette}(2013)}]{Chetrite2013}%
  \BibitemOpen
\bibfield  {journal} {  }\bibfield  {author} {\bibinfo {author} {\bibfnamefont
  {R.}~\bibnamefont {Chetrite}}\ and\ \bibinfo {author} {\bibfnamefont
  {H.}~\bibnamefont {Touchette}},\ }\href@noop {} {\bibfield  {journal}
  {\bibinfo  {journal} {Phys. Rev. Lett.}\ }\textbf {\bibinfo {volume} {111}},\
  \bibinfo {pages} {120601} (\bibinfo {year} {2013})}\BibitemShut {NoStop}%
\bibitem [{\citenamefont {Chetrite}\ and\ \citenamefont
  {Touchette}(2015)}]{Chetrite2015}%
  \BibitemOpen
  \bibfield  {author} {\bibinfo {author} {\bibfnamefont {R.}~\bibnamefont
  {Chetrite}}\ and\ \bibinfo {author} {\bibfnamefont {H.}~\bibnamefont
  {Touchette}},\ }\href@noop {} {\bibfield  {journal} {\bibinfo  {journal}
  {Ann. Henri Poincar\'e}\ }\textbf {\bibinfo {volume} {16}},\ \bibinfo {pages}
  {2005} (\bibinfo {year} {2015})}\BibitemShut {NoStop}%
\bibitem [{\citenamefont {Budini}\ \emph {et~al.}(2014)\citenamefont {Budini},
  \citenamefont {Turner},\ and\ \citenamefont {Garrahan}}]{Budini2014}%
  \BibitemOpen
  \bibfield  {author} {\bibinfo {author} {\bibfnamefont {A.~A.}\ \bibnamefont
  {Budini}}, \bibinfo {author} {\bibfnamefont {R.~M.}\ \bibnamefont {Turner}},
  \ and\ \bibinfo {author} {\bibfnamefont {J.~P.}\ \bibnamefont {Garrahan}},\
  }\href@noop {} {\bibfield  {journal} {\bibinfo  {journal} {J. Stat. Mech.}\
  ,\ \bibinfo {pages} {P03012}} (\bibinfo {year} {2014})}\BibitemShut {NoStop}%
\bibitem [{\citenamefont {Kiukas}\ \emph {et~al.}(2015)\citenamefont {Kiukas},
  \citenamefont {Guta}, \citenamefont {Lesanovsky},\ and\ \citenamefont
  {Garrahan}}]{Kiukas2015}%
  \BibitemOpen
  \bibfield  {author} {\bibinfo {author} {\bibfnamefont {J.}~\bibnamefont
  {Kiukas}}, \bibinfo {author} {\bibfnamefont {M.}~\bibnamefont {Guta}},
  \bibinfo {author} {\bibfnamefont {I.}~\bibnamefont {Lesanovsky}}, \ and\
  \bibinfo {author} {\bibfnamefont {J.~P.}\ \bibnamefont {Garrahan}},\
  }\href@noop {} {\bibfield  {journal} {\bibinfo  {journal} {Phys. Rev. E}\
  }\textbf {\bibinfo {volume} {92}},\ \bibinfo {pages} {012132} (\bibinfo
  {year} {2015})}\BibitemShut {NoStop}%
\bibitem [{\citenamefont {Peliti}(2011)}]{Peliti2011}%
  \BibitemOpen
  \bibfield  {author} {\bibinfo {author} {\bibfnamefont {L.}~\bibnamefont
  {Peliti}},\ }\href@noop {} {\emph {\bibinfo {title} {Statistical mechanics in
  a nutshell}}}\ (\bibinfo  {publisher} {Princeton University Press},\ \bibinfo
  {year} {2011})\BibitemShut {NoStop}%
\bibitem [{\citenamefont {Garrahan}\ \emph {et~al.}(2007)\citenamefont
  {Garrahan}, \citenamefont {Jack}, \citenamefont {Lecomte}, \citenamefont
  {Pitard}, \citenamefont {van Duijvendijk},\ and\ \citenamefont {van
  Wijland}}]{Garrahan2007}%
  \BibitemOpen
  \bibfield  {author} {\bibinfo {author} {\bibfnamefont {J.~P.}\ \bibnamefont
  {Garrahan}}, \bibinfo {author} {\bibfnamefont {R.~L.}\ \bibnamefont {Jack}},
  \bibinfo {author} {\bibfnamefont {V.}~\bibnamefont {Lecomte}}, \bibinfo
  {author} {\bibfnamefont {E.}~\bibnamefont {Pitard}}, \bibinfo {author}
  {\bibfnamefont {K.}~\bibnamefont {van Duijvendijk}}, \ and\ \bibinfo {author}
  {\bibfnamefont {F.}~\bibnamefont {van Wijland}},\ }\href@noop {} {\bibfield
  {journal} {\bibinfo  {journal} {Phys. Rev. Lett.}\ }\textbf {\bibinfo
  {volume} {98}},\ \bibinfo {pages} {195702} (\bibinfo {year}
  {2007})}\BibitemShut {NoStop}%
\bibitem [{\citenamefont {Shmueli}\ \emph {et~al.}(2005)\citenamefont
  {Shmueli}, \citenamefont {Minka}, \citenamefont {Kadane}, \citenamefont
  {Borle},\ and\ \citenamefont {Boatwright}}]{Shmueli2005}%
  \BibitemOpen
  \bibfield  {author} {\bibinfo {author} {\bibfnamefont {G.}~\bibnamefont
  {Shmueli}}, \bibinfo {author} {\bibfnamefont {T.~P.}\ \bibnamefont {Minka}},
  \bibinfo {author} {\bibfnamefont {J.~B.}\ \bibnamefont {Kadane}}, \bibinfo
  {author} {\bibfnamefont {S.}~\bibnamefont {Borle}}, \ and\ \bibinfo {author}
  {\bibfnamefont {P.}~\bibnamefont {Boatwright}},\ }\href@noop {} {\bibfield
  {journal} {\bibinfo  {journal} {J. Roy. Stat. Soc. C-App.}\ }\textbf
  {\bibinfo {volume} {54}},\ \bibinfo {pages} {127} (\bibinfo {year}
  {2005})}\BibitemShut {NoStop}%
\bibitem [{\citenamefont {Lebowitz}\ and\ \citenamefont
  {Spohn}(1999)}]{Lebowitz1999}%
  \BibitemOpen
  \bibfield  {author} {\bibinfo {author} {\bibfnamefont {J.~L.}\ \bibnamefont
  {Lebowitz}}\ and\ \bibinfo {author} {\bibfnamefont {H.}~\bibnamefont
  {Spohn}},\ }\href@noop {} {\bibfield  {journal} {\bibinfo  {journal} {J.
  Stat. Phys.}\ }\textbf {\bibinfo {volume} {95}},\ \bibinfo {pages} {333}
  (\bibinfo {year} {1999})}\BibitemShut {NoStop}%
\bibitem [{\citenamefont {Evans}(2004)}]{Evans2004}%
  \BibitemOpen
  \bibfield  {author} {\bibinfo {author} {\bibfnamefont {R.~M.~L.}\
  \bibnamefont {Evans}},\ }\href@noop {} {\bibfield  {journal} {\bibinfo
  {journal} {Phys. Rev. Lett.}\ }\textbf {\bibinfo {volume} {92}},\ \bibinfo
  {pages} {150601} (\bibinfo {year} {2004})}\BibitemShut {NoStop}%
\bibitem [{\citenamefont {Merolle}\ \emph {et~al.}(2005)\citenamefont
  {Merolle}, \citenamefont {Garrahan},\ and\ \citenamefont
  {Chandler}}]{Merolle2005}%
  \BibitemOpen
  \bibfield  {author} {\bibinfo {author} {\bibfnamefont {M.}~\bibnamefont
  {Merolle}}, \bibinfo {author} {\bibfnamefont {J.~P.}\ \bibnamefont
  {Garrahan}}, \ and\ \bibinfo {author} {\bibfnamefont {D.}~\bibnamefont
  {Chandler}},\ }\href@noop {} {\bibfield  {journal} {\bibinfo  {journal}
  {Proc. Natl. Acad. Sci. USA}\ }\textbf {\bibinfo {volume} {102}},\ \bibinfo
  {pages} {10837} (\bibinfo {year} {2005})}\BibitemShut {NoStop}%
\bibitem [{\citenamefont {Appert-Rolland}\ \emph {et~al.}(2008)\citenamefont
  {Appert-Rolland}, \citenamefont {Derrida}, \citenamefont {Lecomte},\ and\
  \citenamefont {van Wijland}}]{Appert2008}%
  \BibitemOpen
  \bibfield  {author} {\bibinfo {author} {\bibfnamefont {C.}~\bibnamefont
  {Appert-Rolland}}, \bibinfo {author} {\bibfnamefont {B.}~\bibnamefont
  {Derrida}}, \bibinfo {author} {\bibfnamefont {V.}~\bibnamefont {Lecomte}}, \
  and\ \bibinfo {author} {\bibfnamefont {F.}~\bibnamefont {van Wijland}},\
  }\href@noop {} {\bibfield  {journal} {\bibinfo  {journal} {Phys. Rev. E}\
  }\textbf {\bibinfo {volume} {78}},\ \bibinfo {pages} {021122} (\bibinfo
  {year} {2008})}\BibitemShut {NoStop}%
\bibitem [{\citenamefont {Garrahan}\ \emph {et~al.}(2009)\citenamefont
  {Garrahan}, \citenamefont {Jack}, \citenamefont {Lecomte}, \citenamefont
  {Pitard}, \citenamefont {van Duijvendijk},\ and\ \citenamefont {van
  Wijland}}]{Garrahan2009}%
  \BibitemOpen
  \bibfield  {author} {\bibinfo {author} {\bibfnamefont {J.~P.}\ \bibnamefont
  {Garrahan}}, \bibinfo {author} {\bibfnamefont {R.~L.}\ \bibnamefont {Jack}},
  \bibinfo {author} {\bibfnamefont {V.}~\bibnamefont {Lecomte}}, \bibinfo
  {author} {\bibfnamefont {E.}~\bibnamefont {Pitard}}, \bibinfo {author}
  {\bibfnamefont {K.}~\bibnamefont {van Duijvendijk}}, \ and\ \bibinfo {author}
  {\bibfnamefont {F.}~\bibnamefont {van Wijland}},\ }\href@noop {} {\bibfield
  {journal} {\bibinfo  {journal} {J. Phys. A}\ }\textbf {\bibinfo {volume}
  {42}},\ \bibinfo {pages} {075007} (\bibinfo {year} {2009})}\BibitemShut
  {NoStop}%
\bibitem [{\citenamefont {Baiesi}\ \emph {et~al.}(2009)\citenamefont {Baiesi},
  \citenamefont {Maes},\ and\ \citenamefont {Wynants}}]{Baiesi2009}%
  \BibitemOpen
  \bibfield  {author} {\bibinfo {author} {\bibfnamefont {M.}~\bibnamefont
  {Baiesi}}, \bibinfo {author} {\bibfnamefont {C.}~\bibnamefont {Maes}}, \ and\
  \bibinfo {author} {\bibfnamefont {B.}~\bibnamefont {Wynants}},\ }\href@noop
  {} {\bibfield  {journal} {\bibinfo  {journal} {Phys. Rev. Lett.}\ }\textbf
  {\bibinfo {volume} {103}},\ \bibinfo {pages} {010602} (\bibinfo {year}
  {2009})}\BibitemShut {NoStop}%
\bibitem [{\citenamefont {Hedges}\ \emph {et~al.}(2009)\citenamefont {Hedges},
  \citenamefont {Jack}, \citenamefont {Garrahan},\ and\ \citenamefont
  {Chandler}}]{Hedges2009}%
  \BibitemOpen
  \bibfield  {author} {\bibinfo {author} {\bibfnamefont {L.~O.}\ \bibnamefont
  {Hedges}}, \bibinfo {author} {\bibfnamefont {R.~L.}\ \bibnamefont {Jack}},
  \bibinfo {author} {\bibfnamefont {J.~P.}\ \bibnamefont {Garrahan}}, \ and\
  \bibinfo {author} {\bibfnamefont {D.}~\bibnamefont {Chandler}},\ }\href@noop
  {} {\bibfield  {journal} {\bibinfo  {journal} {Science}\ }\textbf {\bibinfo
  {volume} {323}},\ \bibinfo {pages} {1309} (\bibinfo {year}
  {2009})}\BibitemShut {NoStop}%
\bibitem [{\citenamefont {Kurchan}()}]{Kurchan2009}%
  \BibitemOpen
  \bibfield  {author} {\bibinfo {author} {\bibfnamefont {J.}~\bibnamefont
  {Kurchan}},\ }\href@noop {} {\bibinfo  {journal} {arXiv:0901.1271}\
  }\BibitemShut {NoStop}%
\bibitem [{\citenamefont {Esposito}\ \emph {et~al.}(2009)\citenamefont
  {Esposito}, \citenamefont {Harbola},\ and\ \citenamefont
  {Mukamel}}]{Esposito2009}%
  \BibitemOpen
\bibfield  {journal} {  }\bibfield  {author} {\bibinfo {author} {\bibfnamefont
  {M.}~\bibnamefont {Esposito}}, \bibinfo {author} {\bibfnamefont
  {U.}~\bibnamefont {Harbola}}, \ and\ \bibinfo {author} {\bibfnamefont
  {S.}~\bibnamefont {Mukamel}},\ }\href@noop {} {\bibfield  {journal} {\bibinfo
   {journal} {Rev. Mod. Phys.}\ }\textbf {\bibinfo {volume} {81}},\ \bibinfo
  {pages} {1665} (\bibinfo {year} {2009})}\BibitemShut {NoStop}%
\bibitem [{\citenamefont {Garrahan}\ and\ \citenamefont
  {Lesanovsky}(2010)}]{Garrahan2010}%
  \BibitemOpen
  \bibfield  {author} {\bibinfo {author} {\bibfnamefont {J.~P.}\ \bibnamefont
  {Garrahan}}\ and\ \bibinfo {author} {\bibfnamefont {I.}~\bibnamefont
  {Lesanovsky}},\ }\href@noop {} {\bibfield  {journal} {\bibinfo  {journal}
  {Phys. Rev. Lett.}\ }\textbf {\bibinfo {volume} {104}},\ \bibinfo {pages}
  {160601} (\bibinfo {year} {2010})}\BibitemShut {NoStop}%
\bibitem [{\citenamefont {Giardina}\ \emph {et~al.}(2011)\citenamefont
  {Giardina}, \citenamefont {Kurchan}, \citenamefont {Lecomte},\ and\
  \citenamefont {Tailleur}}]{Giardina2011}%
  \BibitemOpen
  \bibfield  {author} {\bibinfo {author} {\bibfnamefont {C.}~\bibnamefont
  {Giardina}}, \bibinfo {author} {\bibfnamefont {J.}~\bibnamefont {Kurchan}},
  \bibinfo {author} {\bibfnamefont {V.}~\bibnamefont {Lecomte}}, \ and\
  \bibinfo {author} {\bibfnamefont {J.}~\bibnamefont {Tailleur}},\ }\href@noop
  {} {\bibfield  {journal} {\bibinfo  {journal} {J. Stat. Phys.}\ }\textbf
  {\bibinfo {volume} {145}},\ \bibinfo {pages} {787} (\bibinfo {year}
  {2011})}\BibitemShut {NoStop}%
\bibitem [{\citenamefont {Budini}(2011)}]{Budini2011}%
  \BibitemOpen
  \bibfield  {author} {\bibinfo {author} {\bibfnamefont {A.}~\bibnamefont
  {Budini}},\ }\href@noop {} {\bibfield  {journal} {\bibinfo  {journal} {Phys.
  Rev. E}\ }\textbf {\bibinfo {volume} {84}},\ \bibinfo {pages} {011141}
  (\bibinfo {year} {2011})}\BibitemShut {NoStop}%
\bibitem [{\citenamefont {Speck}\ and\ \citenamefont
  {Garrahan}(2011)}]{Speck2011}%
  \BibitemOpen
  \bibfield  {author} {\bibinfo {author} {\bibfnamefont {T.}~\bibnamefont
  {Speck}}\ and\ \bibinfo {author} {\bibfnamefont {J.~P.}\ \bibnamefont
  {Garrahan}},\ }\href@noop {} {\bibfield  {journal} {\bibinfo  {journal}
  {Euro. Phys. J. B}\ }\textbf {\bibinfo {volume} {79}},\ \bibinfo {pages} {1}
  (\bibinfo {year} {2011})}\BibitemShut {NoStop}%
\bibitem [{\citenamefont {Speck}\ \emph {et~al.}(2012)\citenamefont {Speck},
  \citenamefont {Malins},\ and\ \citenamefont {Royall}}]{Speck2012b}%
  \BibitemOpen
  \bibfield  {author} {\bibinfo {author} {\bibfnamefont {T.}~\bibnamefont
  {Speck}}, \bibinfo {author} {\bibfnamefont {A.}~\bibnamefont {Malins}}, \
  and\ \bibinfo {author} {\bibfnamefont {C.~P.}\ \bibnamefont {Royall}},\
  }\href@noop {} {\bibfield  {journal} {\bibinfo  {journal} {Phys. Rev. Lett.}\
  }\textbf {\bibinfo {volume} {109}},\ \bibinfo {pages} {195703} (\bibinfo
  {year} {2012})}\BibitemShut {NoStop}%
\bibitem [{\citenamefont {Bodineau}\ and\ \citenamefont
  {Toninelli}(2012)}]{Bodineau2012}%
  \BibitemOpen
  \bibfield  {author} {\bibinfo {author} {\bibfnamefont {T.}~\bibnamefont
  {Bodineau}}\ and\ \bibinfo {author} {\bibfnamefont {C.}~\bibnamefont
  {Toninelli}},\ }\href@noop {} {\bibfield  {journal} {\bibinfo  {journal}
  {Comm. Math. Phys.}\ }\textbf {\bibinfo {volume} {311}},\ \bibinfo {pages}
  {357} (\bibinfo {year} {2012})}\BibitemShut {NoStop}%
\bibitem [{\citenamefont {Flindt}\ and\ \citenamefont
  {Garrahan}(2013)}]{Flindt2013}%
  \BibitemOpen
  \bibfield  {author} {\bibinfo {author} {\bibfnamefont {C.}~\bibnamefont
  {Flindt}}\ and\ \bibinfo {author} {\bibfnamefont {J.~P.}\ \bibnamefont
  {Garrahan}},\ }\href@noop {} {\bibfield  {journal} {\bibinfo  {journal}
  {Phys. Rev. Lett.}\ }\textbf {\bibinfo {volume} {110}},\ \bibinfo {pages}
  {050601} (\bibinfo {year} {2013})}\BibitemShut {NoStop}%
\bibitem [{\citenamefont {Weber}\ \emph {et~al.}(2013)\citenamefont {Weber},
  \citenamefont {Jack},\ and\ \citenamefont {Pande}}]{Weber2013}%
  \BibitemOpen
  \bibfield  {author} {\bibinfo {author} {\bibfnamefont {J.~K.}\ \bibnamefont
  {Weber}}, \bibinfo {author} {\bibfnamefont {R.~L.}\ \bibnamefont {Jack}}, \
  and\ \bibinfo {author} {\bibfnamefont {V.~S.}\ \bibnamefont {Pande}},\
  }\href@noop {} {\bibfield  {journal} {\bibinfo  {journal} {J. Am. Chem.
  Soc.}\ }\textbf {\bibinfo {volume} {135}},\ \bibinfo {pages} {5501} (\bibinfo
  {year} {2013})}\BibitemShut {NoStop}%
\bibitem [{\citenamefont {Espigares}\ \emph {et~al.}(2013)\citenamefont
  {Espigares}, \citenamefont {Garrido},\ and\ \citenamefont
  {Hurtado}}]{Espigares2013}%
  \BibitemOpen
  \bibfield  {author} {\bibinfo {author} {\bibfnamefont {C.~P.}\ \bibnamefont
  {Espigares}}, \bibinfo {author} {\bibfnamefont {P.~L.}\ \bibnamefont
  {Garrido}}, \ and\ \bibinfo {author} {\bibfnamefont {P.~I.}\ \bibnamefont
  {Hurtado}},\ }\href@noop {} {\bibfield  {journal} {\bibinfo  {journal} {Phys.
  Rev. E}\ }\textbf {\bibinfo {volume} {87}},\ \bibinfo {pages} {032115}
  (\bibinfo {year} {2013})}\BibitemShut {NoStop}%
\bibitem [{\citenamefont {Mey}\ \emph {et~al.}(2014)\citenamefont {Mey},
  \citenamefont {Geissler},\ and\ \citenamefont {Garrahan}}]{Mey2014}%
  \BibitemOpen
  \bibfield  {author} {\bibinfo {author} {\bibfnamefont {A.~S. J.~S.}\
  \bibnamefont {Mey}}, \bibinfo {author} {\bibfnamefont {P.~L.}\ \bibnamefont
  {Geissler}}, \ and\ \bibinfo {author} {\bibfnamefont {J.~P.}\ \bibnamefont
  {Garrahan}},\ }\href@noop {} {\bibfield  {journal} {\bibinfo  {journal}
  {Phys. Rev. E}\ }\textbf {\bibinfo {volume} {89}},\ \bibinfo {pages} {032109}
  (\bibinfo {year} {2014})}\BibitemShut {NoStop}%
\bibitem [{\citenamefont {Vaikuntanathan}\ \emph {et~al.}(2014)\citenamefont
  {Vaikuntanathan}, \citenamefont {Gingrich},\ and\ \citenamefont
  {Geissler}}]{Vaikuntanathan2014}%
  \BibitemOpen
  \bibfield  {author} {\bibinfo {author} {\bibfnamefont {S.}~\bibnamefont
  {Vaikuntanathan}}, \bibinfo {author} {\bibfnamefont {T.~R.}\ \bibnamefont
  {Gingrich}}, \ and\ \bibinfo {author} {\bibfnamefont {P.~L.}\ \bibnamefont
  {Geissler}},\ }\href@noop {} {\bibfield  {journal} {\bibinfo  {journal}
  {Phys. Rev. E}\ }\textbf {\bibinfo {volume} {89}},\ \bibinfo {pages} {062108}
  (\bibinfo {year} {2014})}\BibitemShut {NoStop}%
\bibitem [{\citenamefont {Weber}\ \emph {et~al.}(2015)\citenamefont {Weber},
  \citenamefont {Shukla},\ and\ \citenamefont {Pande}}]{Weber2015}%
  \BibitemOpen
  \bibfield  {author} {\bibinfo {author} {\bibfnamefont {J.~K.}\ \bibnamefont
  {Weber}}, \bibinfo {author} {\bibfnamefont {D.}~\bibnamefont {Shukla}}, \
  and\ \bibinfo {author} {\bibfnamefont {V.~S.}\ \bibnamefont {Pande}},\
  }\href@noop {} {\bibfield  {journal} {\bibinfo  {journal} {Proc. Natl. Acad.
  Sci. USA}\ }\textbf {\bibinfo {volume} {112}},\ \bibinfo {pages} {10377}
  (\bibinfo {year} {2015})}\BibitemShut {NoStop}%
\bibitem [{\citenamefont {Jack}\ \emph {et~al.}(2015)\citenamefont {Jack},
  \citenamefont {Thompson},\ and\ \citenamefont {Sollich}}]{Jack2015}%
  \BibitemOpen
  \bibfield  {author} {\bibinfo {author} {\bibfnamefont {R.~L.}\ \bibnamefont
  {Jack}}, \bibinfo {author} {\bibfnamefont {I.~R.}\ \bibnamefont {Thompson}},
  \ and\ \bibinfo {author} {\bibfnamefont {P.}~\bibnamefont {Sollich}},\
  }\href@noop {} {\bibfield  {journal} {\bibinfo  {journal} {Phys. Rev. Lett.}\
  }\textbf {\bibinfo {volume} {114}},\ \bibinfo {pages} {060601} (\bibinfo
  {year} {2015})}\BibitemShut {NoStop}%
\bibitem [{\citenamefont {Ueda}\ and\ \citenamefont {Sasa}(2015)}]{Ueda2015}%
  \BibitemOpen
  \bibfield  {author} {\bibinfo {author} {\bibfnamefont {M.}~\bibnamefont
  {Ueda}}\ and\ \bibinfo {author} {\bibfnamefont {S.}~\bibnamefont {Sasa}},\
  }\href@noop {} {\bibfield  {journal} {\bibinfo  {journal} {Phys. Rev. Lett.}\
  }\textbf {\bibinfo {volume} {115}},\ \bibinfo {pages} {080605} (\bibinfo
  {year} {2015})}\BibitemShut {NoStop}%
\bibitem [{\citenamefont {De~Bacco}\ \emph {et~al.}()\citenamefont {De~Bacco},
  \citenamefont {Guggiola}, \citenamefont {K{\"u}hn},\ and\ \citenamefont
  {Paga}}]{DeBacco2015}%
  \BibitemOpen
  \bibfield  {author} {\bibinfo {author} {\bibfnamefont {C.}~\bibnamefont
  {De~Bacco}}, \bibinfo {author} {\bibfnamefont {A.}~\bibnamefont {Guggiola}},
  \bibinfo {author} {\bibfnamefont {R.}~\bibnamefont {K{\"u}hn}}, \ and\
  \bibinfo {author} {\bibfnamefont {P.}~\bibnamefont {Paga}},\ }\href@noop {}
  {\bibinfo  {journal} {arXiv:1506.08436}\ }\BibitemShut {NoStop}%
\bibitem [{\citenamefont {Szavits-Nossan}\ and\ \citenamefont
  {Evans}(2015)}]{Szavits2015}%
  \BibitemOpen
\bibfield  {journal} {  }\bibfield  {author} {\bibinfo {author} {\bibfnamefont
  {J.}~\bibnamefont {Szavits-Nossan}}\ and\ \bibinfo {author} {\bibfnamefont
  {M.~R.}\ \bibnamefont {Evans}},\ }\href@noop {} {\bibfield  {journal}
  {\bibinfo  {journal} {J. Stat. Mech.}\ ,\ \bibinfo {pages} {P12008}}
  (\bibinfo {year} {2015})}\BibitemShut {NoStop}%
\bibitem [{\citenamefont {Jack}\ and\ \citenamefont
  {Sollich}(2015)}]{Jack2015b}%
  \BibitemOpen
  \bibfield  {author} {\bibinfo {author} {\bibfnamefont {R.~L.}\ \bibnamefont
  {Jack}}\ and\ \bibinfo {author} {\bibfnamefont {P.}~\bibnamefont {Sollich}},\
  }\href@noop {} {\bibfield  {journal} {\bibinfo  {journal} {Eur. Phys.
  J.-Spec. Top.}\ }\textbf {\bibinfo {volume} {224}},\ \bibinfo {pages} {2351}
  (\bibinfo {year} {2015})}\BibitemShut {NoStop}%
\bibitem [{\citenamefont {van Horssen}\ and\ \citenamefont {Gu{\c t}{\u
  a}}(2015)}]{Horssen2015}%
  \BibitemOpen
  \bibfield  {author} {\bibinfo {author} {\bibfnamefont {M.}~\bibnamefont {van
  Horssen}}\ and\ \bibinfo {author} {\bibfnamefont {M.}~\bibnamefont {Gu{\c
  t}{\u a}}},\ }\href@noop {} {\bibfield  {journal} {\bibinfo  {journal} {J.
  Math. Phys.}\ }\textbf {\bibinfo {volume} {56}},\ \bibinfo {pages} {022109}
  (\bibinfo {year} {2015})}\BibitemShut {NoStop}%
\bibitem [{\citenamefont {Pigeon}\ \emph {et~al.}(2015)\citenamefont {Pigeon},
  \citenamefont {Fusco}, \citenamefont {Xuereb}, \citenamefont {De~Chiara},\
  and\ \citenamefont {Paternostro}}]{Pigeon2015}%
  \BibitemOpen
  \bibfield  {author} {\bibinfo {author} {\bibfnamefont {S.}~\bibnamefont
  {Pigeon}}, \bibinfo {author} {\bibfnamefont {L.}~\bibnamefont {Fusco}},
  \bibinfo {author} {\bibfnamefont {A.}~\bibnamefont {Xuereb}}, \bibinfo
  {author} {\bibfnamefont {G.}~\bibnamefont {De~Chiara}}, \ and\ \bibinfo
  {author} {\bibfnamefont {M.}~\bibnamefont {Paternostro}},\ }\href {\doibase
  10.1103/PhysRevA.92.013844} {\bibfield  {journal} {\bibinfo  {journal} {Phys.
  Rev. A}\ }\textbf {\bibinfo {volume} {92}},\ \bibinfo {pages} {013844}
  (\bibinfo {year} {2015})}\BibitemShut {NoStop}%
\bibitem [{\citenamefont {Verley}(2016)}]{Verley2016}%
  \BibitemOpen
  \bibfield  {author} {\bibinfo {author} {\bibfnamefont {G.}~\bibnamefont
  {Verley}},\ }\href@noop {} {\bibfield  {journal} {\bibinfo  {journal} {Phys.
  Rev. E}\ }\textbf {\bibinfo {volume} {93}},\ \bibinfo {pages} {012111}
  (\bibinfo {year} {2016})}\BibitemShut {NoStop}%
\bibitem [{\citenamefont {Bonanca}\ and\ \citenamefont
  {Jarzynski}(2016)}]{Bonanca2016}%
  \BibitemOpen
  \bibfield  {author} {\bibinfo {author} {\bibfnamefont {M.~V.}\ \bibnamefont
  {Bonanca}}\ and\ \bibinfo {author} {\bibfnamefont {C.}~\bibnamefont
  {Jarzynski}},\ }\href@noop {} {\bibfield  {journal} {\bibinfo  {journal}
  {Phys. Rev. E}\ }\textbf {\bibinfo {volume} {93}},\ \bibinfo {pages} {022101}
  (\bibinfo {year} {2016})}\BibitemShut {NoStop}%
\bibitem [{\citenamefont {Nemoto}\ \emph {et~al.}(2016)\citenamefont {Nemoto},
  \citenamefont {Bouchet}, \citenamefont {Jack},\ and\ \citenamefont
  {Lecomte}}]{Nemoto2016}%
  \BibitemOpen
  \bibfield  {author} {\bibinfo {author} {\bibfnamefont {T.}~\bibnamefont
  {Nemoto}}, \bibinfo {author} {\bibfnamefont {F.}~\bibnamefont {Bouchet}},
  \bibinfo {author} {\bibfnamefont {R.~L.}\ \bibnamefont {Jack}}, \ and\
  \bibinfo {author} {\bibfnamefont {V.}~\bibnamefont {Lecomte}},\ }\href@noop
  {} {\bibfield  {journal} {\bibinfo  {journal} {Phys. Rev. E}\ }\textbf
  {\bibinfo {volume} {93}},\ \bibinfo {pages} {062123} (\bibinfo {year}
  {2016})}\BibitemShut {NoStop}%
\bibitem [{\citenamefont {Speck}()}]{Speck2016}%
  \BibitemOpen
  \bibfield  {author} {\bibinfo {author} {\bibfnamefont {T.}~\bibnamefont
  {Speck}},\ }\href@noop {} {\bibinfo  {journal} {arXiv:1601.03540}\
  }\BibitemShut {NoStop}%
\bibitem [{\citenamefont {Jack}\ and\ \citenamefont {Evans}()}]{Jack2016}%
  \BibitemOpen
\bibfield  {journal} {  }\bibfield  {author} {\bibinfo {author} {\bibfnamefont
  {R.~L.}\ \bibnamefont {Jack}}\ and\ \bibinfo {author} {\bibfnamefont
  {R.}~\bibnamefont {Evans}},\ }\href@noop {} {\bibinfo  {journal}
  {arXiv:1602.03815}\ }\BibitemShut {NoStop}%
\bibitem [{\citenamefont {Gherardini}\ \emph {et~al.}(2016)\citenamefont
  {Gherardini}, \citenamefont {Gupta}, \citenamefont {Cataliotti},
  \citenamefont {Smerzi}, \citenamefont {Caruso},\ and\ \citenamefont
  {Ruffo}}]{Gherardini2016}%
  \BibitemOpen
\bibfield  {journal} {  }\bibfield  {author} {\bibinfo {author} {\bibfnamefont
  {S.}~\bibnamefont {Gherardini}}, \bibinfo {author} {\bibfnamefont
  {S.}~\bibnamefont {Gupta}}, \bibinfo {author} {\bibfnamefont {F.~S.}\
  \bibnamefont {Cataliotti}}, \bibinfo {author} {\bibfnamefont
  {A.}~\bibnamefont {Smerzi}}, \bibinfo {author} {\bibfnamefont
  {F.}~\bibnamefont {Caruso}}, \ and\ \bibinfo {author} {\bibfnamefont
  {S.}~\bibnamefont {Ruffo}},\ }\href@noop {} {\bibfield  {journal} {\bibinfo
  {journal} {New J. Phys.}\ }\textbf {\bibinfo {volume} {18}},\ \bibinfo
  {pages} {013048} (\bibinfo {year} {2016})}\BibitemShut {NoStop}%
\bibitem [{\citenamefont {Deffner}(2016)}]{Deffner2016}%
  \BibitemOpen
  \bibfield  {author} {\bibinfo {author} {\bibfnamefont {S.}~\bibnamefont
  {Deffner}},\ }\href@noop {} {\bibfield  {journal} {\bibinfo  {journal} {New
  J. Phys.}\ }\textbf {\bibinfo {volume} {18}},\ \bibinfo {pages} {011001}
  (\bibinfo {year} {2016})}\BibitemShut {NoStop}%
\bibitem [{\citenamefont {Karevski}\ and\ \citenamefont
  {Sch{\"u}tz}()}]{Karevski2016}%
  \BibitemOpen
  \bibfield  {author} {\bibinfo {author} {\bibfnamefont {D.}~\bibnamefont
  {Karevski}}\ and\ \bibinfo {author} {\bibfnamefont {G.}~\bibnamefont
  {Sch{\"u}tz}},\ }\href@noop {} {\bibinfo  {journal} {arXiv:1606.04248}\
  }\BibitemShut {NoStop}%
\bibitem [{\citenamefont {Gardiner}(2004)}]{Gardiner2004}%
  \BibitemOpen
\bibfield  {journal} {  }\bibfield  {author} {\bibinfo {author} {\bibfnamefont
  {C.}~\bibnamefont {Gardiner}},\ }\href@noop {} {\emph {\bibinfo {title}
  {Handbook of stochastic methods}}}\ (\bibinfo  {publisher} {Berlin:
  Springer},\ \bibinfo {year} {2004})\BibitemShut {NoStop}%
\bibitem [{\citenamefont {Bolhuis}(2008)}]{Bolhuis2008}%
  \BibitemOpen
  \bibfield  {author} {\bibinfo {author} {\bibfnamefont {P.~G.}\ \bibnamefont
  {Bolhuis}},\ }\href@noop {} {\bibfield  {journal} {\bibinfo  {journal} {J.
  Chem. Phys.}\ }\textbf {\bibinfo {volume} {129}},\ \bibinfo {pages} {114108}
  (\bibinfo {year} {2008})}\BibitemShut {NoStop}%
\bibitem [{\citenamefont {Harris}\ and\ \citenamefont
  {Touchette}()}]{Harris2016}%
  \BibitemOpen
  \bibfield  {author} {\bibinfo {author} {\bibfnamefont {R.~J.}\ \bibnamefont
  {Harris}}\ and\ \bibinfo {author} {\bibfnamefont {H.}~\bibnamefont
  {Touchette}},\ }\href@noop {} {\bibinfo  {journal} {arXiv:1610.08842}\
  }\BibitemShut {NoStop}%
\bibitem [{\citenamefont {Garrahan}(2016)}]{Garrahan2016}%
  \BibitemOpen
\bibfield  {journal} {  }\bibfield  {author} {\bibinfo {author} {\bibfnamefont
  {J.~P.}\ \bibnamefont {Garrahan}},\ }\href@noop {} {\bibfield  {journal}
  {\bibinfo  {journal} {J. Stat. Mech.}\ ,\ \bibinfo {pages} {073208}}
  (\bibinfo {year} {2016})}\BibitemShut {NoStop}%
\end{thebibliography}%

\end{document}